\documentclass[epj]{svjour}
\usepackage{graphics}
\begin{document}
\title{Transformation from the nonautonomous to standard NLS equations}
\author{Dun Zhao\inst{1,2} \and Xu-Gang He\inst{1} \and Hong-Gang Luo\inst{2,3,4}
}                     % Do not remove
\institute{
  School of Mathematics and Statistics, Lanzhou University, Lanzhou 730000,
  China \and
  Center for Interdisciplinary Studies, Lanzhou University, Lanzhou 730000,
  China \and
  Key Laboratory for Magnetism and Magnetic Materials of the Ministry of Education, Lanzhou University, Lanzhou 730000, China \and
  Institute of Theoretical Physics, Chinese Academy of Sciences, Beijing 100080, China}
\date{Received: date / Revised version: date}

\abstract{In this paper we show a systematical method to obtain
exact solutions of the nonautonomous nonlinear Schr\"odinger (NLS)
equation. An integrable condition is first obtained by the
Painlev\'e analysis, which is shown to be consistent with that
obtained by the Lax pair method. Under this condition, we present
a general transformation, which can directly convert all allowed
exact solutions of the standard NLS equation into the
corresponding exact solutions of the nonautonomous NLS equation.
The method is quite powerful since the standard NLS equation has
been well studied in the past decades and its exact solutions are
vast in the literature. The result provides an effective way to
control the soliton dynamics. Finally, the fundamental bright and
dark solitons are taken as examples to demonstrate its explicit
applications. 
}

\maketitle
\section{Introduction} \label{intro}
The standard nonlinear Schr\"odinger (NLS) equation
\begin{equation}
i\frac{\partial}{\partial T} Q(X, T) + \varepsilon
\frac{\partial^2}{\partial X^2}Q(X, T) + \delta |Q(X, T)|^2Q(X, T)
= 0, \label{std-nls}
\end{equation}
where $\varepsilon$ and $\delta$ are constants, is a fundamental
nonlinear equation to govern system dynamics in many different
fields such as Bose-Einstein condensates
(BEC)\cite{Burger1999,Strecker2002,Becker2008} and nonlinear
optics\cite{Hasegawa1973,Mollenauer1980}. The nature of Eq.
(\ref{std-nls}) has been extensively explored in past decades by
many different ways and its exact solutions including soliton
\cite{ZK1965} are vast in the literature. For a review, one can
refer to Ref. \cite{Sulem1999}. When applied to different
contexts, Eq. (\ref{std-nls}) has many different extensions. For
example, in BEC an additional harmonic external potential is
needed, the resulted equation is well known as Gross-Pitaevskii
equation. In addition, the concept of soliton management has been
extensively explored in recent years\cite{Malomed2006}. The goal
is to control effectively the soliton dynamics. In BEC, the
nonlinear interaction can been easily tuned by an external
magnetic field, namely the Feshbach resonance
management\cite{Kevrekidis2003,Pitaevskii2003}. On the other hand,
in the context of optical soliton communication, the dispersion
management has been explored extensively to improve the optical
soliton communication \cite{hasegawa1995,Nakazawa2000}. In both
cases, the basic equation to describe the system dynamics should
be a generalized nonautonomous NLS equation \cite{Serkin2007},
which reads in one-dimensional(1D) case
\begin{eqnarray}
&& i\frac{\partial u(x,t)}{\partial t} + \varepsilon\,
f(x,t)\frac{\partial ^2
u(x,t)}{\partial x^2} + \delta\, g(x,t)|u(x,t)|^2 u(x,t) \nonumber \\
&& \hspace{4cm} + V(t)x^2 u(x,t) = 0. \label{non-eq1}
\end{eqnarray}
Here $\varepsilon$ and $\delta$ are the same as those in Eq.
(\ref{std-nls}) and $f(x, t)$ and $g(x, t)$ are dimensionless
control parameters of dispersion and nonlinear interaction,
respectively. $V(t)$ is time-dependent harmonic trap potential in
BEC, which is absent in optical transmission line. These
coefficients are assumed usually to be real.

Equation (\ref{non-eq1}) and/or its similar versions are very
difficult to solve because of the time- and space-dependent
dispersion and nonlinear interaction managements and the presence
of the external potential in BEC. Some special solutions have been
obtained by, for example, the Lax pair method
\cite{Serkin2007,Serkin2000,Serkin2002,Serkin2004,liang2005}, the
similarity transformation
\cite{Kruglov2003,Kruglov2004,Kotonop2006,Ponomarenko2007}, and so
on. However, a general method to find solutions of Eq.
(\ref{non-eq1}) has not yet been obtained. Here we obtain a {\it
general} transformation, which can convert all allowed exact
solutions of the standard NLS equation (\ref{std-nls}) to the
corresponding solutions of Eq. (\ref{non-eq1}). To our knowledge,
the result is reported for the first time in the literature, which
provides a straightforward and systematical way to find the exact
solution of the generalized NLS equation (\ref{non-eq1}), as shown
by two examples given in the final part of the paper. Below we
first show this transformation is consistent with the Painlev\'e
integrability condition of Eq. (\ref{non-eq1}).
\section{The Painlev\'e analysis and Transformation} \label{sec:1}
Motivated by the relationship between complete integrability and
the Painlev\'e property of partial differential equations
\cite{Weiss1983,Ablowitz1991}, we perform the WTC test
\cite{Weiss1983} to study possible integrability condition of Eq.
(\ref{non-eq1}). Following the Kruskal ansatz \cite{Jimbo1982} and
the standard procedure of the Painlev\'e analysis
\cite{Ablowitz1991}, we obtain a compatibility condition
\begin{equation}
\frac{g_{t,t}}{g} - \frac{2g_t^2}{g^2} + \frac{f_t^2}{f^2} -
\frac{f_{t,t}}{f} + \frac{g_t}{g}\frac{f_t}{f} + 4\varepsilon f V
= 0, \label{integrability}
\end{equation}
where the subscripts denote the time derivatives. Here we should
mention that the Painlev\'e test requires that $f(x, t)$ and $g(x,
t)$ must be space-independent, i.e.,  $f(x, t) = f(t)$ and $g(x,
t) = g(t)$. It is interesting to note that this condition is
completely consistent with the integrability condition obtained by
Lax pair \cite{Serkin2007}.

The complete integrability of Eq. (\ref{non-eq1}) under the
compatibility condition can also be further confirmed through a
transformation that reduces Eq. (\ref{non-eq1}) to the standard
NLS equation (\ref{std-nls}). Below we look for such a
transformation in a general form of \cite{Zhao2008}
\begin{equation}
u(x,t) = Q(X(x,t),T(t))e^{ia(x,t) + c(t)}, \label{trans1}
\end{equation}
where $X(x,t), T(t), a(x,t)$ and $c(t)$ are real functions to be
determined by the requirement that $u(x,t)$ and $Q(X,T)$ are the
solutions of Eqs. (\ref{non-eq1}) and (\ref{std-nls}),
respectively. Inserting Eq. (\ref{trans1}) into Eq.
(\ref{non-eq1}) and comparing with Eq. (\ref{std-nls}), we obtain
a set of differential equations, which have solutions under the
condition Eq. (\ref{integrability}),
\begin{eqnarray}
&& a(x,t) = \frac{1}{4\,\varepsilon f(t)}\left(\frac{d}{d\,t}\ln
\frac{f(t)}{g(t)}\right)x^2 + C_1 \frac {g(t)}{f(t)}\,x \nonumber \\
&& \hspace{2.5cm} - C_1^2 \varepsilon\, \int \,
\frac {g(t')^2}{f(t')} {d\,t'} + C_2,\label{trans2}\\
&& X(x,t)= \frac { g(t)}{f(t)} x - 2\,C_1\, \varepsilon \int \,
\frac{g(t')^2}{f(t')}{dt'}. \label{trans3} \\
&& T(t) = \int \frac{g^2(t')}{
f(t')}\,d\,t' + C_3, \label{trans4} \\
&& c(t) = \frac{1}{2}\ln \frac{g(t)}{f(t)}, \label{trans5}
\end{eqnarray}
where $C_1, C_2$, and $C_3$ are constants related to the special
boundary conditions and the initial state. Here for simplicity,
they are set to be zero in the following discussions.

The Painlev\'e integrability condition Eq. (\ref{integrability})
is, in fact, a subtle balance condition to keep the nonautonomous
systems integrable. From the management viewpoint of the
solitons\cite{Malomed2006}, Eq. (\ref{integrability}) also
provides an effectively way to manipulate the soliton dynamics.
While any two parameters among $f(t), g(t)$, and $V(t)$ are set,
the remaining one can be tuned according to Eq.
(\ref{integrability}) in order to control the coherent dynamics of
solitons. The applications of Eq. (\ref{integrability}) have been
extensively explored in Ref. \cite{Serkin2007}. However, the
transformation Eqs. (\ref{trans2}) - (\ref{trans5}) have not been
figured out by the Lax pair method. Such transformations are quite
systematic in obtaining the exact solutions of the nonautonomous
NLS equation. For a given nonautonomous NLS equation, we first
check if the coefficients satisfy the compatibility condition Eq.
(\ref{integrability}). If it is true, then the nonautonomous NLS
equation can be reduced to the standard NLS equation
(\ref{std-nls}). All allowed exact solutions, including the
canonical solitons, of the standard NLS equation (\ref{std-nls})
can thus be converted into the corresponding solutions of the
nonautonomous NLS equation. In this sense, a canonical soliton can
be viewed as a ``seed" of the corresponding soliton-like solutions
of Eq. (\ref{non-eq1}) under the compatibility condition Eq.
(\ref{integrability}).

Some remarks are in order. i) If $f(t) = g(t)$ and $V(t) = 0$, the
nonautonomous NLS equation Eq. (\ref{non-eq1}) have the canonical
soliton solutions (up to a phase) regardless of the explicit form
of the time-dependent nonlinearity and dispersion. This is because
in this case the balance between nonlinearity and dispersion is
kept. In this sense the soliton-like solution of Eq.
(\ref{non-eq1}) is a quasi-canonical soliton. ii) When $g(t) \neq
f(t)$, the original balance between nonlinearity and dispersion is
broken down. In this case the canonical soliton must deform itself
to build new balance between nonlinearity and dispersion. In this
sense, the soliton-like solution of Eq. (\ref{non-eq1}) is a
deformed canonical soliton. The amplitude of the soliton will be
scaled by the factor of $\sqrt{g(t)/f(t)}$, as shown by $c(t)$.
This clearly indicates the influence of the dispersion and
nonlinear managements to the soliton behavior. iii) It is very
interesting to note that the confining harmonic external potential
is absent in the transformation equations. However, the presence
of the potential affects the balance between nonlinearity and
dispersion and builds a deep connection between the optical
solitons and the matter-wave ones. iv) If $V(t) = 0$, the solitons
can be quasi-canonical or deformed depending on if $f(t)$ is equal
to $g(t)$ or not, as mentioned above. On the contrary, if $V(t)
\neq 0$, Eq. (\ref{integrability}) indicates that $f(t) \neq
g(t)$. This means that the amplitude of the soliton must change
because of Eq. (\ref{trans5}). This leads to an important
observation that there does not exist the canonical and even
quasi-canonical matter-wave solitons under compatibility condition
Eq. (\ref{integrability}).

It is also helpful to mention some techniques to find the
soliton-like solutions of the nonautonomous NLS equation in the
literature. The Lax pair analysis is very useful in discussing
integrability conditions \cite{Serkin2000,Serkin2007,liang2005}. A
widely used method is the similarity transformation
\cite{Kruglov2003,Kotonop2006,Ponomarenko2007}, which introduces
some explicit transformation parameters. These parameters are
determined by a set of ordinary differential equations, which in
general case can not be solved analytically, as emphasized in Ref.
\cite{Kotonop2006}. Another similarity transformation reducing the
nonautomonous NLS equation to a stationary NLS one has also been
introduced \cite{belmonte-beitia2008}. \textbf{Alternatively, by
the Lie point symmetry group analysis \cite{Gagnon1993}, Eq.
(\ref{non-eq1}) or its similar version can be classified into
different classes and each class can be converted into the
corresponding representative equation by some allowed
transformations. As a result, the exact solutions of the
representative equation can be transformed into the corresponding
solutions of the equations in the same class. However, it was also
pointed out in \cite{Gagnon1993} that in most cases it is
difficult to obtain the exact solutions of these representative
equations and the integrability of certain representative
equations is unclear. Quite different from these techniques, the
present work focuses on the integrability of Eq. (\ref{non-eq1})
and builds a deep connection between the nonautonomous NLS
equation and its autonomous counterpart, which provides a more
systematical way to find solutions of the nonautonomous NLS
equation. Moreover, the corresponding transformation formulas are
explicit and straightforward.} In addition, from the control
viewpoint, our method also provides an effective way to control
the soliton dynamics, as mentioned above.

\section{Applications} Although the transformation
obtained can be applied to all allowed exact solutions of the
standard NLS equation, the further discussion is purposely
restricted to the fundamental bright and dark soliton solutions of
the nonautonomous NLS equation without the harmonic external
potential, which is enough to show how the soliton dynamics is
controlled by corresponding tunable parameters. In this case, Eq.
(\ref{integrability}) becomes
\begin{equation}
f(t) = g(t) \exp\left(-\alpha \int g(t')\, d\, t'\right),
\label{fg1}
\end{equation}
where $\alpha$ is a constant and the transformation equations
(\ref{trans2})-(\ref{trans5}) become $$a(x,t) = -\frac{\alpha}{4}
\exp(G_{\alpha}(t)) x^2,$$
$$X(x,t) = \exp(G_{\alpha}(t)) x,$$
$$T(t) = \int dt'\,g(t') \exp(G_{\alpha}(t'))$$, and
$$c(t) = (1/2) G_{\alpha}(t),$$ where $G_{\alpha}(t) = \alpha \int^t_0
g(t')\,d\,t'$.

\begin{figure}
\resizebox{0.5\textwidth}{!}{
  \includegraphics{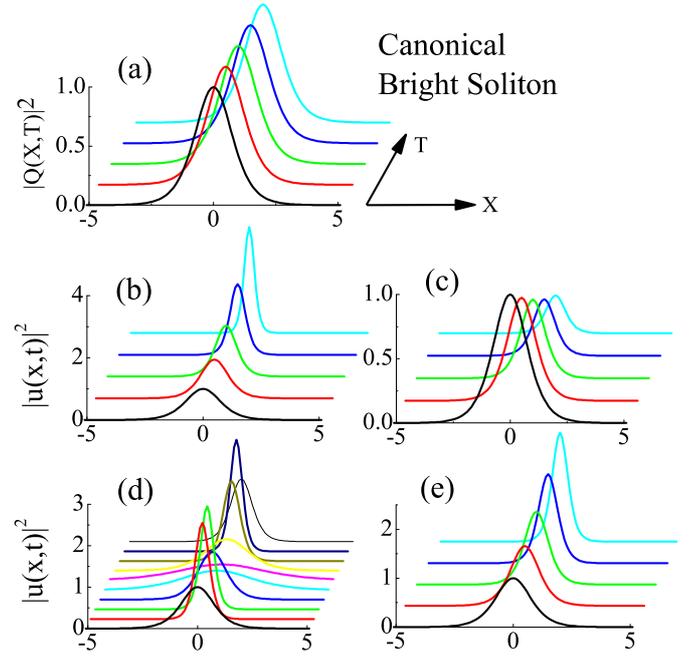}
}
\caption{(a) The canonical bright soliton of Eq. (\ref{std-nls})
with $\varepsilon = 1/2$ and $\delta = 1$ and the corresponding
bright soliton-like solutions of Eq. (\ref{non-eq1}) for different
nonlinearity modulations: (b) $g(t) = \exp(t)$, (c) $g(t) =
\exp(-t)$, (d) $g(t) = \cos(t)$, and (e) $g(t) = 1$. In all cases
$\alpha = 1$. } \label{fig1}
\end{figure}
\begin{figure}
\resizebox{0.5\textwidth}{!}{
  \includegraphics{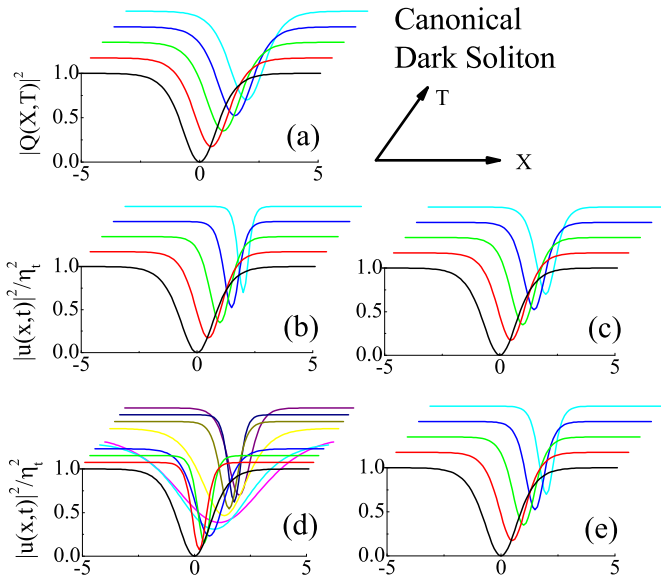}
}
\caption{The canonical dark soliton of Eq. (\ref{std-nls}) with
$\varepsilon = -1/2$ and $\delta = 1$ and the corresponding dark
soliton-like solutions of Eq. (\ref{non-eq1}) for different
nonlinearity modulations same as those in Fig. \ref{fig2}. For
clarity, the amplitude of the dark soliton-like solutions is
normalized by $\eta_t = \exp\left(\frac{1}{2}G_1(t)\right)$.}
\label{fig2}
\end{figure}

When $\varepsilon = 1/2$ and $\delta = 1$ ($\delta \varepsilon >
0$), Eq. (\ref{std-nls}) has the fundamental canonical bright
soliton solution $Q(X, T) = \mbox {sech}(X)\exp(iT/2)$ and when
$\varepsilon = -1/2$ and $\delta = 1$ ($\delta \varepsilon < 0$),
the fundamental dark soliton solution of Eq. (\ref{std-nls}) has
the form of $Q(X, T) = \tanh(X)\exp(iT)$. Starting from these two
solutions, we show the corresponding soliton solutions of Eq.
(\ref{non-eq1}) for four different cases: $g(t) = 1, \exp(t),
\exp(-t)$, and $\cos(t)$, which represent constant, enhancement,
suppression, and periodic modulations of nonlinearity,
respectively. We emphasize that these modulations can be readily
realized, for example, by the Feshbach resonance technique in the
Bose-Einstein condensate context. The corresponding dispersion
modulations follow Eq. (\ref{fg1}). It is noted that $\alpha = 0$
leads to $G_0(t) = 0$, which is trivial up to a phase, as
mentioned above. A nonzero $\alpha$ has nontrivial results and
without loss of generality we take $\alpha = 1$ below. In Fig.
\ref{fig1} and Fig. \ref{fig2} we explicitly present the bright
soliton-like and the dark soliton-like solutions for four
different nonlinearity modulations, respectively. For comparison,
we also plot the canonical solitons, as shown in Fig.
\ref{fig1}(a) and Fig. \ref{fig2}(a), respectively.

Fig. \ref{fig1}(b) and (c) show that the canonical bright soliton
becomes more and more either sharper or broader depending on
enhancement or suppression of the nonlinearity. When the
nonlinearity keeps unchanged and the dispersion is suppressed, the
canonical bright soliton also becomes more and more sharper, as
shown in Fig. \ref{fig1}(e). This can be understood by the fact
that the soliton is due to the balance between dispersion and
nonlinearity. Most interesting case is Fig. \ref{fig1}(d), where
the nonlinearity modulation is periodic. As a result, the
canonical bright soliton is also modulated periodically. All these
results indicate that the bright soliton-like solution and its
canonical counterpart has a close relationship. For different
nonlinearity modulations, we have checked the integration of $\int
|u(x, t)|^2 dx$ and found it keeps unchange in time, which further
shows the nature of the bright soliton-like solutions of Eq.
(\ref{non-eq1}). The similar result is also true to the dark
soliton-like solutions, as shown in Fig. \ref{fig2}. These results
shed light on the understanding of the soliton dynamics and
provide an exact way to make a dispersion and/or nonlinearity
management of solitons. It is expected to have a realistic
application to the optical soliton communication technologies and
the matter-wave soliton dynamics.

Finally, it should be pointed out that the present analysis can
also be applied to all exact solutions of Eq. (\ref{std-nls}),
including the multi-soliton cases. This provide a systematical way
to study the dynamics of the nonautonomous NLS equation.

\section{Conclusion} We propose a systematical way to find
the exact solutions of the nonautonomous NLS equation. The
nonautonomous NLS system obtained are completely integrable and
the soliton-like solutions result from a balance between
dispersion, nonlinearity, and/or an external potential applied,
just like the canonical soliton. This result builds a unified
picture of the nonautonomous and canonical NLS equations and
provides an effective way to control the soliton dynamics.

Support from the NSFC and the Program for NCET of China is
acknowledged.

\end{document}